\documentclass{IEEEtran}

\usepackage{amsmath,amsfonts,amssymb,mathrsfs}
\usepackage{graphicx}
\usepackage[sort]{cite}        
\usepackage{enumerate}
\usepackage{wrapfig}
\usepackage{subcaption}
\usepackage{color}

\newtheorem{assume}{Assumption}[section]
\newtheorem{theorem}{Theorem}[section]

\newtheorem{lemma}{Lemma}[section]

\newtheorem{definition}{Definition}[section]
\newtheorem{proposition}{Proposition}[section]

\begin{document}
		\title{Safe Stabilization using Nonsmooth Control Lyapunov Barrier Function}
		
		\author{Jianglin Lan$^1$, Eldert van Henten$^2$, Peter Groot Koerkamp$^2$, and Congcong Sun$^2$
			
		\thanks{This work was supported by the  
		Leverhulme Trust Early Career Fellowship under Award ECF-2021-517. }
		
		\thanks{$^1$Jianglin Lan is with the James Watt School of Engineering, University of Glasgow, Glasgow G12 8QQ, U.K.
		(email: Jianglin.Lan@glasgow.ac.uk)}
		
		\thanks{$^2$Eldert van Henten, Peter Groot Koerkamp, and Congcong Sun are with the Agricultural Biosystems Engineering Group, Wageningen University \& Research, 6700 AC Wageningen, The Netherlands
			(emails: \{eldert.vanhenten, peter.grootkoerkamp, congcong.sun\}@wur.nl)}

	}
	
	\maketitle

\begin{abstract}
	This paper addresses the challenge of safe stabilization, ensuring the system state reach the origin while avoiding unsafe regions. Existing approaches relying on smooth Lyapunov barrier functions often fail to guarantee a feasible controller. To overcome this limitation, we introduce the nonsmooth Control Lyapunov Barrier Function (NCLBF), which ensures the existence of a safe and stabilizing controller. We provide a systematic framework for designing NCLBF and feedback control strategies to achieve safe stabilization in the presence of multiple bounded unsafe regions. Theoretical analysis and simulations of both linear and nonlinear systems demonstrate the effectiveness and superiority of our approach compared to the existing smooth functions method.
\end{abstract}

	
\section{Introduction}\label{sec:intro}
Stabilization, driving the system state to the origin, is essential in control systems, while safety is critical for applications like aircraft and autonomous vehicles. Thus, a feedback controller must ensure \textit{safe stabilization}—maintaining stability while avoiding unsafe states—aligning with the “stable while avoid” framework \cite{edwards2023general}.

A control Lyapunov function (CLF) ensures asymptotic controllability to the origin \cite{artstein1983stabilization}, leading most studies to design smooth CLFs for continuous feedback stabilization, such as Sontag's universal control law \cite{sontag1989universal}.
Recently, control barrier functions (CBFs) have been widely used to ensure safety by enforcing the forward invariance of system trajectories \cite{Ames+19c, dawson2022safe, Schneeberger+24}.

A common approach to ensuring safe stabilization is incorporating separately constructed CLF and CBFs into an optimization problem to minimally adjust a nominal controller \cite{ames2016control,li2023survey}. While flexible, this approach can lead to infeasibility due to conflicts among CLF and CBFs \cite{mestres2024feasibility,wang2024safe}.
An alternative is unifying CLF and CBF into a single function. The control Lyapunov barrier function (CLBF) \cite{romdlony2014uniting,romdlony2016stabilization} achieves this via a linear combination, ensuring a closed-form feedback controller using Sontag's universal control law. CLBFs have been integrated with model predictive control \cite{wu2019control,zheng2024control}, sliding mode control \cite{gomez2022notion}, learning-based control \cite{du2023reinforcement,dawson2022safe}. However, CLBFs can introduce undesired local equilibria due to gradient cancellation between CLF and CBF \cite{braun2019complete}. Additionally, smooth CLBFs fail to exist in systems with bounded unsafe sets, preventing continuous feedback control for safe stabilization \cite{braun2020comment}.
To address this, \cite{wang2020stabilization} modifies CLBFs by treating undesired stationary points as unsafe states, but this relies on identifying these points, which is challenging for complex nonlinear systems. Another approach, converse Lyapunov barrier functions \cite{meng2022smooth}, lacks constructive theorems.
Thus, smooth CLF-CBF unification is limited, motivating the use of nonsmooth functions for broader applicability in safe stabilization.

Nonsmooth CLFs are used to construct non-continuous stabilization controllers for nonlinear and switched systems \cite{sontag1995nonsmooth, liberzon2003switching}. Nonsmooth CBFs, using Boolean composition with nondifferentiable max and min operators, ensure safety under multiple unsafe sets \cite{glotfelter2017nonsmooth}. The complete control Lyapunov function (CCLF) \cite{braun2019complete} attempts to unify CLF and CBF but is limited to linear systems with a single unsafe state point. The Dini derivative is used to evaluate the convergence of CCLF \cite{braun2019complete}, but computing it is challenging \cite{cortes2024mandalay}, creating a bottleneck for nonsmooth functions. 

This paper proposes a novel nonsmooth function to unify CLF and CBF for safe stabilization. The main contributions are as follows:
\begin{itemize}
	\item We introduce the nonsmooth control Lyapunov barrier function (NCLBF) with a systematic construction approach. Unlike CLBF \cite{romdlony2014uniting,romdlony2016stabilization}, NCLBF prevents gradient cancellation between CLF and CBF and remains valid with bounded unsafe sets, ensuring broader applicability. Compared to CCLF \cite{braun2019complete}, NCLBF extends to multiple unsafe state sets.  
	\item The NCLBF is utilized to design a piecewise continuous feedback controller for safe stabilization subject to multiple unsafe sets. Compared to the CCLF-based method \cite{braun2019complete}, our control design is applicable to both linear and Lipschitz control-affine nonlinear systems.
\end{itemize}

The paper is structured as follows: Section \ref{sec:problemsetup} describes the problem and relevant concepts, Section \ref{sec:single} presents the NCLBF-based control for single unsafe set, Section \ref{sec:multiple} extends the design to multiple unsafe sets, Section \ref{sec:example} presents simulations, and Section \ref{sec:conclusion} concludes.

\textit{Notations:} 
The symbol $\mathbb{R}^n$ denotes the $n$-dimensional Euclidean space and $\mathbb{R}_{\geq 0}$ denotes the set of non-negative real numbers.
For $x, y \in \mathbb{R}^n$, $\| x \|$ denotes the 2-norm, $\langle x,y \rangle = x^\top y$, $\mu(y) = y^\top / \| y \|^2$, and $\bar{\mu}(y) = [\mu(y_1), \cdots, \mu(y_n)]^\top$.
$\mathcal{O}$ denotes an open set, with closure $\overline{\mathcal{O}}$. $\partial \mathcal{S}$ represents the boundary of set $\mathcal{S}$. 
For a function $h(x): \mathcal{D} \mapsto \mathbb{R}, \mathcal{D} \subseteq \mathbb{R}^n$, its gradient is $\nabla h(x)$ if continuously differentiable, with time derivative $\dot{h}(x) = \nabla h(x) \dot{x}$; otherwise, its generalized gradient set is $\partial h(x)$. 
A continuous function $\rho: \mathbb{R}_{\geq 0} \mapsto \mathbb{R}_{\geq 0}$ is class-$\mathcal{P}$ if $\rho(0) = 0$, $\rho(s) > 0, \forall s > 0$.
A function $\alpha \in \mathcal{P}$ is class-$\mathcal{K}$ if it is strictly increasing and class-$\mathcal{K}_\infty$ if $\lim_{s \rightarrow \infty} \alpha(s) = \infty$.
A function $\rho: \mathbb{R}_{\geq 0} \mapsto \mathbb{R}_{\geq 0}$ is class-$\mathcal{L}$ if it is strictly decreasing and $\lim_{s \rightarrow \infty} \alpha(s) = 0$.
A function $\beta: \mathbb{R}_{\geq 0}^2 \mapsto \mathbb{R}_{\geq 0}$ is
class-$\mathcal{KL}$ if $\beta(\cdot, s) \in \mathcal{K}_\infty, \forall s \in \mathbb{R}_{\geq 0}$ and $\beta(s, \cdot) \in \mathcal{L}, \forall s \in \mathbb{R}_{\geq 0}$.

\section{Problem description and preliminaries}\label{sec:problemsetup}

Consider a class of continuous-time affine-control system
\begin{equation}\label{eq:sys1}
	\dot{x} = f(x) + g(x) u 
\end{equation}
with the state $x \in \mathbb{R}^{n}$, control input $u \in \mathbb{R}^{m}$, and nonlinear functions $f(x) \in \mathbb{R}^{n}$ and $g(x) \in \mathbb{R}^{n \times m}$. Let $\mathcal{X} \subset \mathbb{R}^n$ be the operational state space and $\mathcal{F}(x)$ denote the set of solutions to \eqref{eq:sys1} starting at $x(0)$. 
The functions $f(x)$ and $g(x)$ are locally Lipschitz continuous, $f(0) = 0$, $g(x)$ has full rank $m$, and the origin $x = 0$ is the desired equilibrium of \eqref{eq:sys1}. To simplify notations, we will express \eqref{eq:sys1} as $\dot{x} = F(x,u)$ or $\dot{x} = F(x)$ under a state-feedback controller $u = \kappa(x)$. 


This work aims to design a feedback controller to ensure safe stabilization of system \eqref{eq:sys1} in the sense of $\mathcal{KL}$-stability and safety \cite{braun2021stability}, as in Definition \ref{defn:stability and safe}.

\begin{definition}\label{defn:stability and safe}
Consider the system $\dot{x} = F(x,u)$ with the state space $\mathcal{X} \subset \mathbb{R}^n$ and disjoint unsafe state sets $\mathcal{O}_i \subset \mathcal{X}$, $i \in [1,N]$, satisfying $0 \notin \mathcal{O} := \cup_{i=1}^N \mathcal{O}_i$. 
This system is said to be $\mathcal{KL}$-stable w.r.t. the origin and never enters $\mathcal{O}$ if there exists a feedback controller $u = \kappa(x)$ such that $\forall x(0) \in \mathcal{X} \setminus \mathcal{O}$, 
\begin{equation}
	\| x(t) \| \leq \beta(\|x(0) \|, t), ~ x(t) \notin  \mathcal{O}, ~ \forall t \geq 0,
\end{equation}
where $\beta$ is a $\mathcal{KL}$ function.
\end{definition}

According to Definition \ref{defn:stability and safe}, $\lim_{\| x(0) \| \rightarrow 0}\beta(\| x(0)\|, t) = 0$ and
$\lim_{t \rightarrow \infty} \beta(\| x(0)\|, t) = 0$. Thus, for finite $x(0)$, $\| x(t) \|$ remains bounded for all $t \geq 0$ and is asymptotically stable, i.e. $\lim_{t \rightarrow \infty} \| x(t) \| = 0$.
The control design in this paper uses a nonsmooth control Lyapunov barrier function (NCLBF) based on a generalized derivative for convergence analysis, adapted from the Mandalay derivative \cite{cortes2024mandalay} (Definition \ref{defn:upper deriv}).

\begin{definition}\label{defn:upper deriv}
Let $h(x,u): \mathcal{D} \mapsto \mathbb{R}$, $\mathcal{D} \subseteq \mathbb{R}^n \times \mathbb{R}^m$, be a locally Lipschitz continuous function  and $\mathcal{F}: \mathbb{R}^n \times \mathbb{R}^m \mapsto \mathbb{R}^n$ be a set-valued map. 
	The upper generalized derivative $\overline{D}_{\mathcal{F}} h(x,u)$ and lower generalized derivative $\underline{D}_{\mathcal{F}} h(x,u)$ of $h$ w.r.t. $\mathcal{F}$ at $(x,u)$ are defined as
	\begin{equation}
		\overline{D}_{\mathcal{F}} h(x,u) = \sup L_{\mathcal{F}} h(x,u), ~ \underline{D}_{\mathcal{F}} h(x,u) = \inf L_{\mathcal{F}} h(x,u),	\nonumber    
	\end{equation}
	with the set-valued Lie derivative $L_{\mathcal{F}} h(x,u) = \{\langle \xi, v \rangle \in \mathbb{R} \mid v \in \mathcal{F}(x,u), \xi \in \partial h(x,u) \}$.
	When $\overline{D}_{\mathcal{F}} h(x,u) = \underline{D}_{\mathcal{F}} h(x,u)$, $h$ is generalized differentiable with the generalized derivative $D_{\mathcal{F}} h(x,u) = \overline{D}_{\mathcal{F}} h(x,u) = \underline{D}_{\mathcal{F}} h(x,u)$.
\end{definition}

As shown in \cite{cortes2024mandalay}, when $h$ is locally Lipschitz and $\mathcal{F}$ is compact, both $\overline{D}_{\mathcal{F}} h(x,u)$ and $\underline{D}_{\mathcal{F}} h(x,u)$ are finite.
If $\mathcal{F}$ is a singleton and $h$ is continuously differentiable, $D_{\mathcal{F}} h(x,u)$ reduces to the usual derivative. 
For simplicity, we denote $\mathcal{F}(x,u)$ as $\mathcal{F}(x)$ and $D_{\mathcal{F}} h(x,u)$ as $D_{\mathcal{F}} h(x)$.

The Dini derivative is commonly used to evaluate the convergence of nonsmooth functions like CBFs \cite{glotfelter2017nonsmooth} and CCLF \cite{braun2019complete}, but computing them is challenging. In contrast, as illustrated in \cite{cortes2024mandalay}, $\overline{D}_\mathcal{F} h(x,u)$ and $\underline{D}_{\mathcal{F}} h(x,u)$ can be computed by expressing $V(x)$ as a composite function with simpler components, making it easier to implement. 

The proposed safe stabilization control leverages the concept of NCLBF (Definition \ref{defn:NCLBF}) for unifying CLF and CBFs.
\begin{definition}\label{defn:NCLBF}
	Consider the system $\dot{x} = F(x)$ with the state space $\mathcal{X} \subset \mathbb{R}^n$, desired equilibrium $x_g$, solution set $\mathcal{F}$, and disjoint unsafe state sets $\mathcal{O}_i \subset \mathcal{X}$, $i \in [1,N]$, satisfying $x_g \notin \mathcal{O} := \cup_{i=1}^N \mathcal{O}_i$.
	A Lipschitz continuous function $V(x): \mathbb{R}^n \mapsto \mathbb{R}$ is called an NCLBF, if there exist functions $\alpha_1, \alpha_2 \in \mathcal{K}_\infty$ and $\rho \in \mathcal{P}$, and constants $c_i > 0, i \in [1,N]$, such that 
	\begin{subequations}\label{NCLBF conds}
		\begin{align}
			\label{NCLBF cond1}
			& V(x) = c_i, \forall x \in \partial \mathcal{O}_i, ~ c_i \leq \min_{x \in \mathcal{O}_i} V(x), ~ i \in [1,N], \\
			\label{NCLBF cond2}
			& \alpha_1(\| x - x_g \|) \leq V(x) \leq \alpha_2(\| x - x_g \|), ~ \forall x \in \mathcal{X}, \\
			\label{NCLBF cond3}
			& \overline{D}_\mathcal{F} V(x) \leq -\rho(\| x \|), ~ \forall x \in \mathcal{X} \setminus (\mathcal{O} \cup \{x_g\}). 
		\end{align}
	\end{subequations} 	
\end{definition}

Condition \eqref{NCLBF cond1} indicates that the level-set $V(x) = c_i$ separates the state trajectory from each unsafe set $\mathcal{O}_i$, $i \in [1,N]$. 
Condition \eqref{NCLBF cond2} ensures that $V(x)$ is positive definite except at $x = x_g$, radially unbounded, and $V(x_g) = 0$. 
Under \eqref{NCLBF cond1} and \eqref{NCLBF cond2}, $V(x)$ carries the properties of CBFs and a CLF, respectively. 
Furthermore, \eqref{NCLBF cond3} ensures that $V(x)$ is decreasing for all $ x \in \mathcal{X} \setminus (\mathcal{O} \cup \{x_g\})$ and thus converges to $V(x_g) = 0$. 

The NCLBF concept is closely related to CCLF \cite[Definition 3]{braun2019complete} but generalizes to an arbitrary equilibrium $x_g$ and replaces the hard-to-compute Dini derivative with the upper generalized derivative in condition \eqref{NCLBF cond3}. 
We use $\overline{D}_\mathcal{F} V(x)$ instead of $D_\mathcal{F} V(x)$ because the former always exists, whereas the latter exists only when  $\overline{D}_\mathcal{F} V(x) = \underline{D}_\mathcal{F} V(x)$ (Definition \ref{defn:upper deriv}). This ensures that \eqref{NCLBF cond3} always implies $D_\mathcal{F} V(x) \leq -\rho(\| x \|), ~ \forall x \in \mathcal{X} \setminus (\mathcal{O} \cup x_g)$.

\begin{lemma}\label{thm:NCLBF}
	For system \eqref{eq:sys1} with disjoint unsafe state sets $\mathcal{O}_i \subset \mathcal{X}$, $i \in [1,N]$, where $0 \notin \mathcal{O} := \cup_{i=1}^N \mathcal{O}_i$, if $V(x)$ is an NCLBF under the state-feedback controller $u = \kappa(x)$ and $x(0) \in \mathcal{X} \setminus \mathcal{O}$, then the controller ensures safe stabilization.
\end{lemma}

The proof of Lemma \ref{thm:NCLBF} follows with minor changes from \cite[Theorem 2]{braun2019complete} and is omitted. Using this lemma, Sections \ref{sec:single} and \ref{sec:multiple} outline a systematic framework for designing NCLBF and feedback control for safe stabilization.

\section{NCLBF-based control for a single unsafe set}\label{sec:single}
\subsection{Construction of NCLBF}
Consider that the unsafe states are enclosed by an open ball
\begin{equation}\label{NCLBF:obstacle}
\mathcal{O} = \{ x \in \mathcal{X} \setminus \{0\} \mid \| x - x_c \| < \sqrt{r} \},
\end{equation}
where $\mathcal{X}$ is the state space, $x_c$ is the centre, and $\sqrt{r}$ is the radius. 
The $n$-ball may yield a conservative control law for safe stabilization but can always enclose a bounded unsafe region of arbitrary shape (excluding the equilibrium 0). Exploring less conservative representations is left for future work. 

We define two functions:
\begin{equation}\label{defn Lx, Bx}
	L(x) = \| x \|^2,~ B(x) = \eta_2 - \eta_1 \| x - x_c \|^2,
\end{equation}
where $\eta_1$ and $\eta_2$ are design constants, 
and three state sets:
\begin{align}\label{thm3:3 regions}
	\begin{split}
		\mathcal{R}_1 &= \{ x \in \mathcal{X} \mid B(x) > L(x),  x \notin \mathcal{O} \}, \\
		\mathcal{R}_2 &= \{ x \in \mathcal{X} \mid B(x) < L(x) \}, \\
		\mathcal{R}_3 &= \{ x \in \mathcal{X} \mid B(x) = L(x)\}. 
	\end{split}		
\end{align} 
The set $\mathcal{R}_3$ serves as a virtual boundary enclosing the ball 
 $\mathcal{O}$ and $x(0)$ is required to be outside $\mathcal{R}_3$ in this paper. 

Theorem \ref{thm3:construct NCLBF} provides a method for constructing an NCLBF.

\begin{theorem}\label{thm3:construct NCLBF}
For system $\dot{x} = F(x)$ with the state space $\mathcal{X}$, equilibrium 0, and an unsafe state set $\mathcal{O}$ in \eqref{NCLBF:obstacle}, it is $\mathcal{KL}$-stable and safe for all 
$x(0) \in \mathcal{X} \setminus (\mathcal{O} \cup \mathcal{R}_1 \cup \mathcal{R}_3)$ if
\begin{enumerate}
	\item[1)] The NCLBF $V(x)$ is constructed as
	\begin{align}\label{thm3:NCLBF}
		V(x) = \max (L(x), B(x))		
	\end{align}
	with $L(x)$ and $B(x)$ in \eqref{defn Lx, Bx} and $\eta_1$ and $\eta_2$ satisfying 
	\begin{align}\label{thm3:eta}
		\begin{split}
		\eta_1 \geq (\|x_c\| + \sqrt{r}) / (\|x_c\| - \sqrt{r}), \\
		\eta_1 r + \max_{x \in \overline{\mathcal{O}}} L(x) \leq \eta_2 < \eta_1 \|x_c\|^2.
		\end{split}
	\end{align}
	\item[2)] There exists a function $\rho(\| x \|) \in \mathcal{P}$ such that
	\begin{align} \label{thm3:generalized derivative}	
		\overline{D}_{\mathcal{F}} V(x) \leq -\rho(\| x \|), 
	\end{align}
	with $\overline{D}_{\mathcal{F}} V(x)$ in the form of
	\begin{align} \label{thm3:generalized derivative2}
		\overline{D}_{\mathcal{F}} V(x) = d_i(x), ~ x \in \mathcal{R}_i, ~ i \in [1,3],
	\end{align}
	where 
	$d_1(x) = \nabla B \cdot F(x)$, $d_2(x) = \nabla L \cdot F(x)$, $d_3(x) = 0.5 (d_1(x) + d_2(x)) + 0.5 |d_1(x) - d_2(x)|$, and $\mathcal{R}_i, i \in [1,3]$, are in \eqref{thm3:3 regions}.
\end{enumerate}
\end{theorem}
\begin{IEEEproof}
For all $x \in \mathcal{O}$, $0 \leq \| x - x_c \|^2 < r$ leading to
$\eta_2 - \eta_1 r < B(x) \leq \eta_2$. 
Since $\eta_2 - \eta_1 r \geq \max_{x \in \overline{\mathcal{O}}} L(x)$, it follows that $B(x) > L(x), \forall x \in \mathcal{O}$, implying
$V(x) = B(x). $
Thus, $V(x)$ satisfies the condition \eqref{NCLBF cond1} with $c = \eta_2 - \eta_1 r > 0$.
Finally, to ensure 0 is outside the region $B(x) \leq L(x)$, we use $\eta_2 < \eta_1 \|x_c\|^2$, yielding $B(0) < L(0)$.
Since $V(x)$ follows \eqref{thm3:NCLBF}, \eqref{NCLBF cond2} holds with $\mathcal{K}_\infty$ functions 
$\alpha_1(\|x\|) = \| x \|^2$ and $\alpha_2(\|x\|) = \| x \|^2 + \max_{\| y \| \leq \|x\|} V(y)$.

To verify condition \eqref{NCLBF cond3}, rearranging $V(x)$ as
$$V(x) = h_0(x) + h_1(h_2(x)),$$ 
where $h_0(x) = 0.5 (B(x) + L(x))$, $h_1(h_2(x)) = 0.5 |h_2(x)|$, and $h_2(x) = B(x) - L(x)$. Both $h_0(x)$ and $h_2(x)$ are continuously differentiable, while $h_1(h_2)$ is non-differentiable at $h_2(x) = 0$ but locally Lipschitz in $\mathcal{X}$. Thus, we derive the generalized derivative of $V(x)$ in the three regions of \eqref{thm3:3 regions}. 

\textit{1) Region $\mathcal{R}_1$}: 
In this case, $h_2(x) > 0$ and $h_1(h_2(x)) = 0.5 h_2(x) = 0.5 (B(x) - L(x))$, so $V(x) = B(x)$, which is continuously differentiable with the generalized derivative 
$D_\mathcal{F} V(x) = \nabla B \cdot F(x)$.

\textit{2) Region $\mathcal{R}_2$}: 
In this case, $h_2(x) < 0$ and $h_1(h_2(x)) = 0.5(L(x) - B(x))$, so $V(x) = L(x)$, which is continuously differentiable with 
$D_\mathcal{F} V(x) = \nabla L \cdot F(x)$. 

\textit{3) Region $\mathcal{R}_3$}: 
In this case, $h_2(x) = 0$ and $V(x)$ is non-differentiable. 
At the points where $h_2(x) = 0$, we derive
\begin{align}
	\partial h_1 &= 
	\begin{cases}
		-0.5, & h_2(x) \rightarrow 0^{-} \\
		0.5, & h_2(x) \rightarrow 0^{+} 
	\end{cases}, \nonumber\\
	\partial h_2 &= \nabla B \cdot - \nabla L \cdot, ~
	\partial h_0 = 0.5 (\nabla B \cdot + \nabla L \cdot). \nonumber
\end{align} 
Since $h_2(x)$ is differentiable, $D_\mathcal{F} h_2(x) = \dot{h}_2(x)$ and it follows that
$D_\mathcal{F} (h_1(h_2))(x) \leq  0.5 |\dot{h}_2(x)|$ and thus
\begin{align}
		D_\mathcal{F} V(x) 
	={}& \dot{h}_0(x) + D_\mathcal{F} (h_1(h_2))(x) \nonumber\\
	\leq{}&  \dot{h}_0(x) + 0.5 |\dot{h}_2(x)| \nonumber\\
	={}&
	0.5 (\nabla B \cdot + \nabla L \cdot) F(x) + 0.5 |(\nabla B \cdot - \nabla L \cdot) F(x)|. \nonumber
\end{align} 
Combining the three cases leads to the upper generalized derivative of $V(x)$ in \eqref{thm3:generalized derivative2}. 
Hence, if there exists a function $\rho(\| x \|) \in \mathcal{P}$ such that $\overline{D}_{\mathcal{F}} V(x) \leq -\rho(\| x \|)$ in all the three state regions,
then the condition \eqref{NCLBF cond3} is satisfied. 

Since $0 \in \mathcal{R}_2$ and $V(x) = L(x)$ in $\mathcal{R}_2$, we have $V(0) = 0$ and $D_\mathcal{F} V(x) = 0$, meaning $x$ remaining at 0. Thus, condition \eqref{NCLBF cond3} holds for all $x \in \mathcal{X} \setminus \mathcal{O}$. In summary, the NCLBF $V(x)$ satisfies all conditions in Definition \ref{defn:NCLBF}, so by Lemma \ref{thm:NCLBF}, 
the system is safe and $\mathcal{KL}$-stable for all $x(0) \in \mathcal{X} \setminus \mathcal{O}$. 
\end{IEEEproof}

The smooth CLBF \cite{romdlony2016stabilization} linearly combines CLF and CBF, which can lead to unwanted local equilibria due to gradient cancellation. In contrast, the proposed NCLBF \eqref{thm3:NCLBF} uses the $\max$ operation, ensuring that 0 is the strict global minimum. 

\subsection{Properties of NCLBF}\label{subsec:properties NCLBF}
\begin{figure}[t]
	\centering
	\includegraphics[width=0.6\linewidth]{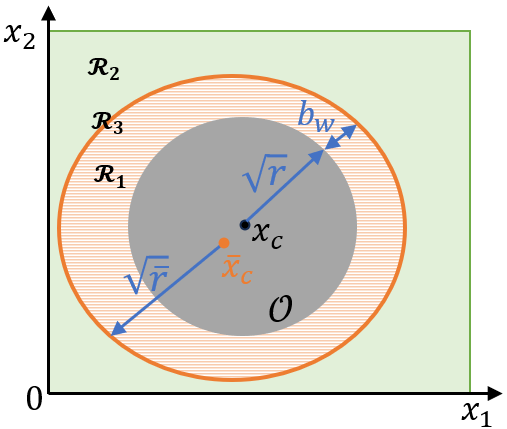}
	\vspace{-1mm}
	\caption{A 2-D illustration of the proposed NCLBF design. The state regions $\mathcal{R}_1$, $\mathcal{R}_2$ and $\mathcal{R}_3$ are defined in \eqref{thm3:3 regions}. The unsafe boundary $\partial \mathcal{O}$ has a radius $\sqrt{r}$ centred at $x_c$, while $\mathcal{R}_3$ has a radius $\sqrt{\bar{r}}$ centred at $\bar{x}_c$. The buffer width $b_w$ is the shortest distance between these boundaries.}
	\label{fig1}
\end{figure}

Figure \ref{fig1} provides a 2-D illustration of the proposed NCLBF. 

\textit{1) Buffer width.}
For the proposed NCLBF \eqref{thm3:NCLBF}, the points satisfying $x \in \mathcal{R}_3$ lie on the sphere $\mathcal{S}$ defined by
\begin{equation}\label{sphere:Bx=Lx}
	\| x - \bar{x}_c \|^2 = \bar{r}	
\end{equation}
with the centre $\bar{x}_c = \eta_1 x_c / (1+\eta_1)$ and radius $\sqrt{\bar{r}}$ satisfying $\bar{r} = [(1+\eta_1) \eta_2 - \eta_1 \| x_c \|^2 ] / (1+\eta_1)^2$. 
Since $\mathcal{O}$ is an $n$-ball, $\max_{x \in \overline{\mathcal{O}}} L(x) = (\| x_c \| + \sqrt{r})^2$.
We set $\eta_2$ as
\begin{equation}\label{sphere1:Bx=Lx3}
	\eta_2 = \eta_1 r + (\| x_c \| + \sqrt{r})^2 + w	
\end{equation}
with $0 < w < \eta_1 (\|x_c\|^2 - r) - (\|x_c\| + \sqrt{r})^2$. 

The distance between the centres $x_c$ and $\bar{x}_c$ is $\| x_c - \bar{x}_c \| = \|x_c\| / (1 + \eta_1)$. Thus, we can write $\bar{r}$ as
\begin{equation}\label{sphere2:Bx=Lx3}
\bar{r} = ( \sqrt{r} + \|x_c\| / (1 + \eta_1) )^2 + b_w^2
\end{equation}
with the buffer width $b_w = \sqrt{w / (\eta_1 + 1)}$, representing the shortest distance between $\mathcal{R}_3$ and $\partial \mathcal{O}$. We can adjust $\eta_1$ and $w$ to tune this width.

\begin{figure}[t]
	\centering
	\includegraphics[width=0.8\linewidth]{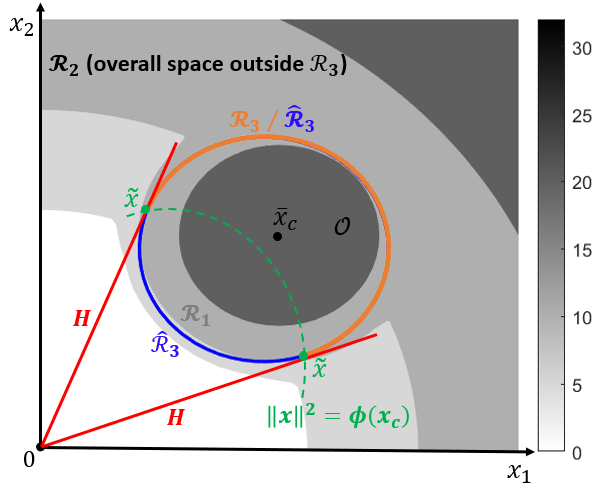}
	\vspace{-1mm}
	\caption{A simple 2-D illustration shows the shrinking of the region (the sphere $\|x - \bar{x}_c\|^2 = \bar{r}$) by removing $\hat{\mathcal{R}}_3 = \left\{ x \in \mathcal{R}_3 \mid \|x\|^2 < \phi(x_c) \right\}$. The greyscale (darker indicates higher values) represents the NCLBF $V(x)$ in the state space $\mathcal{X} := \mathcal{R}_2 \cup \mathcal{R}_3 \cup \mathcal{R}_1 \cup \mathcal{O}$. 
		The state regions $\mathcal{R}_1$, $\mathcal{R}_2$ and $\mathcal{R}_3$ are defined in \eqref{thm3:3 regions}, and $H$ represents a tangent plane at $\tilde{x}$.}
	\label{fig2}
\end{figure}

\textit{2) Shrunk state region $\mathcal{R}_3$.}
From \eqref{thm3:generalized derivative2}, $\overline{D}_{\mathcal{F}} V(x)$ in $\mathcal{R}_3$ involves the absolute operator, complicating the evaluation of the decreasing condition \eqref{thm3:generalized derivative}, especially for control design. 
Proposition \ref{prop:NCLBF reduce R3} shows that $\mathcal{R}_3$ can be shrunk to simplify evaluation, whose key idea is illustrated in Fig. \ref{fig2}.  

\begin{proposition}\label{prop:NCLBF reduce R3}
For the NCLBF constructed in Theorem \ref{thm3:construct NCLBF}, the state region $\mathcal{R}_3$ can be shrunk to $\mathcal{R}_3 \setminus \hat{\mathcal{R}}_3$, removing 
\begin{equation}\label{region R3 reduced}
\hat{\mathcal{R}}_3 = \left\{ x \in \mathcal{R}_3 \mid \|x\|^2 < \phi(x_c) \right\}, 
\end{equation}
where $\phi(x_c) = (\eta_1 \|x_c\|^2 - \eta_2) / (\eta_1 + 1)$. Moreover, $\mathcal{R}_3$ is not a limit cycle for the state trajectory.
\end{proposition}
\begin{IEEEproof}
For the sphere $\mathcal{S}$ in \eqref{sphere:Bx=Lx}, define $S(x) = \|x - \bar{x}_c\|^2$. Let $H(\tilde{x},\bar{x}_c)$ be a hyperplane tangent to $\mathcal{S}$ at $x = \tilde{x}$ and passing through the origin. The gradient of $S(x)$ at $\tilde{x}$ is $\nabla S(\tilde{x}) = 2 (\tilde{x} - \bar{x}_c)^\top$.
Since the tangent plane $H(\tilde{x},\bar{x}_c)$ is orthogonal to $\nabla S(\tilde{x})$ at $x = \tilde{x}$, it is described by 
\begin{equation}\label{pf2:prop reduce R3}
\nabla S(\tilde{x}) \cdot (x - \tilde{x}) = 0.
\end{equation}
Substituting $x = 0$ and $\nabla S(\tilde{x})$ into \eqref{pf2:prop reduce R3} gives
\begin{equation}\label{pf3:prop reduce R3}
	\|\tilde{x}\|^2 = \tilde{x}^\top \bar{x}_c.
\end{equation}
When $n > 2$, there are infinitely many tangent planes $H(\tilde{x},\bar{x}_c)$ of the sphere $\mathcal{S}$, with corresponding infinite contacting points $\tilde{x}$, all satisfying \eqref{pf3:prop reduce R3}.
The contacting points are defined by 
\begin{equation}\label{pf4:prop reduce R3}
\mathcal{X}_t = \{ x \in \mathcal{R}_3 \mid \|x\|^2 = x^\top \bar{x}_c \}.
\end{equation}
From \eqref{pf4:prop reduce R3}, the contacting points $x$ satisfy $\|x - \bar{x}_c\|^2 = \bar{r}$ and $\|x\|^2 = x^\top \bar{x}_c$. Solving these equations with $\bar{x}_c$ and $\bar{r}$ from \eqref{sphere:Bx=Lx} gives 
$\|x\|^2 = \phi(x_c) = (\eta_1 \|x_c\|^2 - \eta_2) / (\eta_1 + 1)$.

Let $\hat{\mathcal{R}}_3 = \left\{ x \in \mathcal{R}_3 \mid \|x\|^2 < \phi(x_c) \right\}$ and $\mathcal{R}_{2,s}$ be the set of states enclosed by the hyperplanes beneath $\mathcal{R}_3$, satisfying $\mathcal{R}_{2,s} \subset \mathcal{R}_2$. A 2-D graphic view of these state sets is shown in Fig. \ref{fig2}. 
When $x \in \hat{\mathcal{R}}_3$, the hyperplane $H$ and  $\mathcal{R}_{2,s}$ belong to $\mathcal{R}_2$, where $V(x) = L(x) = \|x\|^2$. 
If the decreasing condition in $\mathcal{R}_2$ holds, $L(x) = \|x\|^2$ converges exponentially to 0, causing the level values of $H$ to monotonically decrease towards 0. Thus, for $x \in H \cap \hat{\mathcal{R}}_3$, the state trajectories will converge to 0 and never reach $\hat{\mathcal{R}}_3$ from below for any $x(0) \in \mathcal{X} \setminus (\mathcal{O} \cup \mathcal{R}_1 \cup \mathcal{R}_3)$. 
Additionally, if the decreasing condition holds for $\mathcal{R}_1$ and $\mathcal{R}_3 \setminus \hat{\mathcal{R}}_3$, the state trajectories reaching $\mathcal{R}_3 \setminus \hat{\mathcal{R}}_3$ or entering $\mathcal{R}_1$ will be bounced back to $\mathcal{R}_2$, ensuring the trajectories starting from $x(0) \in \mathcal{X} \setminus (\mathcal{O} \cup \mathcal{R}_1 \cup \mathcal{R}_3)$ never reach $\hat{\mathcal{R}}_3$ from the above. 

In summary, the decreasing condition in $\mathcal{R}_3$ needs to be evaluated only for states where $\|x\|^2 \geq \phi(x_c)$. 
Additionally, even if a state trajectory moves along  $\mathcal{R}_3$ near the origin, it will leave $\mathcal{R}_3$ at the contacting points in \eqref{pf4:prop reduce R3} and converge to the origin, confirming that $\mathcal{R}_3$ is not a limit cycle. 
\end{IEEEproof}

By Proposition \ref{prop:NCLBF reduce R3}, $\hat{\mathcal{R}}_3$ can be ignored when evaluating the decreasing condition \eqref{thm3:generalized derivative}.

\subsection{NCLBF-based Control Design}\label{subsec:control single}
Based on the constructed NCLBF, we first consider the control design for system \eqref{eq:sys1} with a single unsafe set, which requires Assumption \ref{assume:Pg}.

\begin{assume}\label{assume:Pg}
	For all $x \in \{ x \in \mathcal{R}_2 \cup \mathcal{R}_3 \mid \nabla L \cdot g(x) \!=\! 0  \}$, $\nabla L \cdot f(x) \leq 0$.	
	For all $x \in \{ x \in \mathcal{R}_1 \cup \mathcal{R}_3 \mid \nabla B \cdot g(x) = 0  \}$, $\nabla B \cdot f(x) \leq 0$.	
	The system is zero-state detectable w.r.t. $\nabla L \cdot g(x)$ and $\nabla B \cdot g(x)$, i.e., $\nabla L \cdot g(x) = 0 ~ \forall t \geq 0 \implies x(t) \rightarrow 0$ while $\nabla B \cdot g(x) = 0 ~ \forall t \geq 0 \implies x(t) \rightarrow 0$.
\end{assume}

Define the generalized derivative of the NCLBF $V(x)$ along $f(x)$ and $g(x)$ as $D_f V(x)$ and $D_g V(x)$, respectively. Assumption \ref{assume:Pg} is compactly expressed as: 
$D_f V(x) \leq 0 ~ \forall x \in \{ x \in \mathcal{X} \setminus \mathcal{O} \mid D_g V(x) = 0  \}$, and 
$D_g V(x) = 0 ~ \forall t \geq 0 \implies x(t) \rightarrow 0$. 
This ensures the properness of NCLBF, which is also necessary for the CLBF-based control design \cite{romdlony2014uniting,romdlony2016stabilization}. Given $x(0) \in \mathcal{X} \setminus (\mathcal{O} \cup \mathcal{R}_1 \cup \mathcal{R}_3)$, then Assumption \ref{assume:Pg} can be relaxed by shrinking the set $\mathcal{R}_3$ as in Proposition \ref{prop:NCLBF reduce R3}. 

For the control design, the corresponding generalized upper derivative $\overline{D}_{\mathcal{F}} V(x)$ is directly obtained 
by replacing $F(x)$ in \eqref{thm3:generalized derivative2} with $f(x) + g(x) u$.
As shown in the proof of Theorem \ref{thm3:construct NCLBF}, at time step $t$, $\overline{D}_{\mathcal{F}} V(x)$ in $\mathcal{R}_3$ can be represented by
\begin{align}\label{eq1:new R3}
	\begin{split}
		\overline{D}_{\mathcal{F}} V(x(t)) &= 0.5 (d_1(x(t)) + d_2(x(t))) \\
		&\quad + \partial h_1 \cdot (d_1(x(t)) - d_2(x(t))), \\
		\partial h_1 & = 
		\begin{cases}
			-0.5, & h_2(x(t)) \rightarrow 0^{-} \\
			0.5, & h_2(x(t)) \rightarrow 0^{+} 
		\end{cases},
	\end{split}	
\end{align}
where $d_1(x(t)) = \nabla B \cdot (f(x(t)) + g(x(t)) u(t))$, $d_2(x(t)) = \nabla L \cdot (f(x(t)) + g(x(t)) u(t))$, 
and $h_2(x(t)) = B(x(t)) - L(x(t))$.

Let $x(t-t_s)$ be the state at time step $t_s$ backward. Then, $\partial h_1 = 0.5 ~ \mathrm{sign}(h_2(x(t-t_s)))$, where $\mathrm{sign}(\cdot)$ is the signum function. Thus, from \eqref{eq1:new R3}, it follows that
\begin{align}\label{eq2:new R3}
	&\overline{D}_{\mathcal{F}} V(x(t)) = 0.5 \left[ 1 + \mathrm{sign}(h_2(x(t-t_s))) \right] d_1(x(t)) \nonumber\\
	&+ 0.5 \left[ 1 - \mathrm{sign}(h_2(x(t-t_s))) \right] d_2(x(t)).
\end{align}
If $h_2(x(t-t_s)) < 0$ (i.e. $x(t-t_s) \in \mathcal{R}_2$), then $h_2(x(t)) \rightarrow 0^{-}$ and $\mathrm{sign}(h_2(x(t-t_s))) = -1$, meaning the state trajectory approaches $\mathcal{R}_3$ from $\mathcal{R}_2$. If $h_2(x(t-t_s)) > 0$ (i.e. $x(t-t_s) \in \mathcal{R}_1$), then $\mathrm{sign}(h_2(x(t-t_s))) = 1$ and $h_2(x(t)) \rightarrow 0^{+}$, meaning the state trajectory approaches $\mathcal{R}_3$ from $\mathcal{R}_1$. Hence, \eqref{eq2:new R3} is equivalent to
\begin{equation}\label{eq3:new R3}
	\!\!\!\overline{D}_{\mathcal{F}} V(x(t)) \!\!=\!\! 
	\begin{cases}
		d_2(x(t)), \!\!\!&\! (x(t) \!\in\! \mathcal{R}_3) \!\cap\! (x(t-t_s) \!\in\! \mathcal{R}_2) \\
		d_1(x(t)), \!\!\!&\! (x(t) \!\in\! \mathcal{R}_3) \!\cap\! (x(t-t_s) \!\in\! \mathcal{R}_1)
	\end{cases}.\!\!\!\!\!\!
\end{equation}
The proposed controller is described in Theorem \ref{thm:control design}.

\begin{theorem}\label{thm:control design}
	Consider system \eqref{eq:sys1} under Assumption \ref{assume:Pg}, with the state space $\mathcal{X}$ and an unsafe state set $\mathcal{O}$ in \eqref{NCLBF:obstacle}. Using the NCLBF $V(x)$ in \eqref{thm3:NCLBF}, safe stabilization is achieved for $x(0) \in \mathcal{X} \setminus (\mathcal{O} \cup \mathcal{R}_1 \cup \mathcal{R}_3)$ through the controller
	\begin{equation}\label{controller:thm4}
		u = \kappa_i(x), ~x \in \mathcal{R}_i, ~i \in [1,3]
	\end{equation}
	with the sets $\mathcal{R}_i$, $i \in [1,3]$, in \eqref{thm3:3 regions} and the feedback laws
	\begin{subequations}\label{controllers:thm4}
		\begin{align}
			\label{u1:thm4}
			\kappa_1(x) &\!=\! 
			\begin{cases}
				-\mu(B_g) B_f - \mathbf{c}_1 \bar{\mu}(B_g) \| x \|^2, \!\!&\!\! B_g \neq 0 \\
				0, \!\!&\!\! B_g = 0
			\end{cases},\!\!
			\\
			\label{u2:thm4}
			\kappa_2(x) &\!=\! 
			\begin{cases}
				- (L_f \!+\! \sqrt{L_f^2 \!+\! \gamma \| L_g \|^4}) \mu(L_g), \!\!&\!\! L_g \neq 0 \\
				0, \!\!&\!\! L_g = 0	
			\end{cases}, \!\!\\
			\label{u3:thm4}
			\kappa_3(x) &\!=\! 
			\begin{cases}
				\kappa_1(x), \!\!&\!\! (x(t) \in \mathcal{R}_3) \cap (x(t-t_s) \in \mathcal{R}_1) \\
				\kappa_2(x), \!\!&\!\! (x(t) \in \mathcal{R}_3) \cap (x(t-t_s) \in \mathcal{R}_2) \\
				\kappa_2(x), \!\!&\!\! (x(t) \in \mathcal{R}_3) \cap (x(t-t_s) \in \mathcal{R}_3)
			\end{cases},\!\!\!\!
		\end{align}	
	\end{subequations}
	with $B_f = \nabla B \cdot f(x)$, $B_g = \nabla B \cdot g(x)$,
	$L_f = \nabla L \cdot f(x)$, $L_g = \nabla L \cdot g(x)$, the design constants
	$\mathbf{c}_1 = \mathrm{diag}(c_{1,1}, \cdots, c_{1,m}) > 0$, $\gamma > 0$, and $\eta_1$ and $\eta_2$ satisfying \eqref{thm3:eta}. More specifically, $\eta_2$ is designed as
	\begin{equation}\label{eta2:thm4}
		\eta_2 = \eta_1 r + (\| x_c \| + \sqrt{r})^2 + w	
	\end{equation}
	with $0 < w < \eta_1 (\|x_c\|^2 - r) - (\|x_c\| + \sqrt{r})^2$. 
\end{theorem}
\begin{IEEEproof}
	Following Theorem \ref{thm3:construct NCLBF}, the NCLBF $V(x)$ in \eqref{thm3:NCLBF} trivially satisfies \eqref{NCLBF cond1} and \eqref{NCLBF cond2}, due to independence on $u$. We further shows that the controller \eqref{controller:thm4} ensures \eqref{NCLBF cond3} below. 
	
	\textit{1) Region $\mathcal{R}_1$}: 
	Here, $V(x) = B(x)$. 
	Applying $u = \kappa_1(x)$ in \eqref{u1:thm4},  $\overline{D}_{\mathcal{F}} V(x) = \nabla B \cdot (f(x) + g(x) \kappa_1(x) ) = \nabla B \cdot f(x) + B_g \kappa_1(x)$.
	Since $B_g \mu(B_g) B_f = B_f$ and 
	$B_g \mathbf{c}_1 \bar{\mu}(B_g) \| x \|^2 = \bar{c}_1 \| x \|^2$, where $\bar{c}_1 = \sum_{j=1}^{n} c_{1,j} > 0$, $\overline{D}_{\mathcal{F}} V(x)$ becomes
	\begin{equation}\label{pf2:thm4}
		\overline{D}_{\mathcal{F}} V(x) =
		\begin{cases}
			-\bar{c}_1 \| x \|^2, & B_g \neq 0 \\
			B_f, & B_g = 0 
		\end{cases}.
	\end{equation} 
	By Assumption \ref{assume:Pg}, $B_f \leq 0$ when $B_g = 0$. Thus, from \eqref{pf2:thm4}, $\forall x \in \mathcal{R}_1$, $\overline{D}_{\mathcal{F}} V(x) \leq -\rho(\|x\|)$ with the function $\rho(\|x\|) = \rho_0 \| x \|^2 \in \mathcal{P}$ and $\rho_0 > 0$, meaning that
	\eqref{NCLBF cond3} is satisfied. 
	
	\textit{2) Region $\mathcal{R}_2$}: 
	In this case, $V(x) = L(x) = \| x \|^2$. 
	The controller $u = \kappa_2(x)$ in \eqref{u2:thm4} is Sontag's universal control law.
	Since $2 x^\top (f(x) + g(x) \kappa_2(x)) = L_f + L_g \kappa_2(x)$, the corresponding $\overline{D}_{\mathcal{F}} V(x)$ is
	\begin{equation}\label{pf4:thm4}
		\overline{D}_{\mathcal{F}} V(x) =
		\begin{cases}
			-\sqrt{L_f^2 + \gamma \|L_g\|^4}, & L_g \neq 0 \\
			L_f, & L_g = 0 
		\end{cases}.
	\end{equation} 
	By Assumption \ref{assume:Pg}, $L_f \leq 0$ when $L_g = 0$. Hence, from \eqref{pf4:thm4},
	$\forall x \in \mathcal{R}_2 \setminus \{0\}$, $\overline{D}_{\mathcal{F}} V(x) \leq -\rho(\|x\|)$ with a $\mathcal{P}$-class function $\rho(\|x\|) = \rho_0 \| x \|^2$ and $\rho_0 > 0$, meaning that \eqref{NCLBF cond3} is satisfied.
	
	\textit{3) Region $\mathcal{R}_3$}: 
	The design of $u(t) = \kappa_3(x(t))$ in \eqref{u3:thm4} is divided into three disjoint cases based on the previous time step state $x(t-t_s)$. 
	\textit{Case 1:} When $(x(t) \in \mathcal{R}_3) \cap (x(t-t_s) \in \mathcal{R}_1)$, from \eqref{eq3:new R3}, $\overline{D}_F V(x) = \nabla B \cdot (f(x) + g(x) u)$. Designing $\kappa_3(x) = \kappa_1(x)$ ensures \eqref{NCLBF cond3}, as shown in the proof of Region $\mathcal{R}_1$;
	\textit{Case 2:} When $(x(t) \in \mathcal{R}_3) \cap (x(t-t_s) \in \mathcal{R}_2)$, from \eqref{eq3:new R3}, $\overline{D}_F V(x) = d_2(x) = \nabla L \cdot (f(x) + g(x) u)$. Designing $\kappa_3(x) = \kappa_2(x)$ ensures \eqref{NCLBF cond3}, as shown in the proof of Region $\mathcal{R}_2$; 
	\textit{Case 3:}
	When $(x(t) \in \mathcal{R}_3) \cap (x(t-t_s) \in \mathcal{R}_3)$, the trajectory moves along $\mathcal{R}_3$ at time $t$. By designing $\kappa_3(x(t)) = \kappa_2(x(t))$, the state trajectory remains tangent to $\mathcal{R}_3$ and pointing towards the origin. As shown in Proposition \ref{prop:NCLBF reduce R3}, when the trajectory reaches the contacting points in \eqref{pf4:prop reduce R3}, it will leave $\mathcal{R}_3$ and converge to the origin. 
	
	While the controller transforms \eqref{eq:sys1} into a piecewise affine system, the right-hand side, $F(x,u)$, remains bounded on any compact set of $(x,u)$. Moreover, the set of discontinuities of $F(x,u)$ has measure zero, ensuring that \eqref{eq:sys1} admits well-defined solutions in the sense of Filippov \cite{pavlov2007convergence}. Since the same NCLBF, $V(x)$, is used globally in the state space $\mathcal{X}$, safe stabilization is guaranteed by Lemma \ref{thm:NCLBF}. 
\end{IEEEproof}

The way to select the design constants is as follows: 
1) $c_{1,i}, i \in [1,m]$, and $\gamma$ are positive constants. Larger values lead to bigger control actions and faster convergence to equilibrium.
2) $\eta_1$ satisfies \eqref{thm3:eta} and $\eta_2$ is designed using  $\eta_1$ and $w$ from \eqref{eta2:thm4}. The choice of $\eta_1$ and $w$ affects the buffer width $b_w$, balancing state safety and the allowable initial state space.

\section{Handling multiple unsafe sets}\label{sec:multiple}

The method in Section \ref{sec:single} is directly extended to cover multiple unsafe sets. Consider $N$ disjoint unsafe state sets in $\mathcal{X}$, with the $i$-th unsafe set enclosed by 
\begin{equation}\label{multi:obstacles}
	\mathcal{O}_i = \{ x \in \mathcal{X} \setminus \{0\} \mid \| x - x_{i,c} \| < \sqrt{r_i} \},
\end{equation}
where $x_{i,c}$ is the centre and $\sqrt{r_i}$ is the radius. 

We define the functions:
\begin{equation}\label{defn Lx, Bx2}
	L(x) \!=\! \| x \|^2, ~
	B_i(x) \!=\! \eta_{i,2} - \eta_{i,1} \| x - x_{i,c} \|^2, i \in [1,N],
\end{equation}
where $\eta_{i,1}$ and $\eta_{i,2}$ are design constants,
and the state sets:
\begin{align}\label{thmmulti:region}
	\begin{split}
	\mathcal{R}_{i,1} &= \{ x \in \mathcal{X} \mid x \notin (\cup_{i=1}^N \mathcal{O}_i), \\
	& \hspace{1.8em}B_i(x) > \max (L(x), \{B_j(x)\}_{j=1,j\neq i}^N ) \}, \\
	\mathcal{R}_2 &= \{ x \in \mathcal{X} \mid \max_{j \in [1,N]} B_j(x) < L(x) \},\\
	\mathcal{R}_{i,3} &= \{ x \in \mathcal{X} \mid B_i(x) = L(x)\}, \\
	\hat{\mathcal{R}}_{i,3} &= \{ x \in \mathcal{R}_{i,3} \mid \|x\|^2 < \phi(x_{i,c}) \},
	\end{split}
\end{align}
where $\phi(x_{i,c}) = (\eta_{i,1} \|x_{i,c}\|^2 - \eta_{i,2}) / (\eta_{i,1} + 1)$.
Theorem \ref{thmmulti:construct NCLBF} provides a way for constructing a single NCLBF to certify $\mathcal{KL}$-stability and safety in this case.

\begin{theorem}\label{thmmulti:construct NCLBF}
	For system $\dot{x} = F(x)$ with the state space $\mathcal{X}$ and the unsafe state sets $\mathcal{O}_i$, $i \in [1,N]$,	in \eqref{multi:obstacles}, it is $\mathcal{KL}$-stable and safe for  $x(0) \in \mathcal{X} \setminus (\mathcal{O} \cup (\cup_{i=1}^{N} \mathcal{R}_{i,1}) \cup (\cup_{i=1}^{N} \mathcal{R}_{i,3}))$, where $\mathcal{O} = \cup_{i=1}^N \mathcal{O}_i$, if 
	\begin{enumerate}
		\item[1)] The NCLBF $V(x)$ is constructed as
		\begin{align}\label{thmmulti:NCLBF}
			V(x) = \max ( L(x), \underset{i \in [1,N]}{\max} B_i(x) ), 	
		\end{align}
		with $L(x)$ and $B_i(x)$ in \eqref{defn Lx, Bx2}, and	
		\begin{subequations}\label{thmmulti:eta}
			\begin{align}
				\label{thmmulti:eta1}
				& \!\!	 \eta_{i,1} > (\|x_{i,c}\| + \sqrt{r_i}) / (\|x_{i,c}\| - \sqrt{r_i}), \\
				& \!\!\eta_{i,1} r_i + \underset{x \in \overline{\mathcal{O}}_i}{\max} \|x\|^2 \leq \eta_{i,2} < \eta_{i,1} \|x_{i,c}\|^2, \!\!\!\\
				\label{thmmulti:eta2}
				& \!\! \| x_{i,c} - x_{j,c} \| \!>\! \sqrt{\bar{r}_i} \!+\! \sqrt{\bar{r}_j}, \forall i,j \!\in\! [1,N], i \neq j, \!\!\!
			\end{align}		
		\end{subequations}
		where 	
		$\bar{r}_i = \frac{(1+\eta_{i,1}) \eta_{i,2} - \eta_{i,1} \| x_{i,c} \|^2 }{(1+\eta_{i,1})^2}, ~ i \in [1,N]$. 
		\item[2)] There exists a function $\rho(\| x \|) \in \mathcal{P}$ satisfying
		\begin{align} \label{thmmulti:Mandalay}
			\overline{D}_{\mathcal{F}} V(x) \leq -\rho(\| x \|)
		\end{align}
		with $\overline{D}_{\mathcal{F}} V(x)$ in the form of
		\begin{align}
			\overline{D}_{\mathcal{F}} V(x)
			= 
			\begin{cases}
				d_{i,1}(x), & x \in \mathcal{R}_{i,1}, ~ i \in [1,N] \\
				d_2(x), & x \in \mathcal{R}_2 \\
				d_{i,3}(x), &
				x \in \mathcal{R}_{i,3} \setminus \hat{\mathcal{R}}_{i,3}, ~ i \in [1,N]
			\end{cases} \nonumber
		\end{align}
		where $d_{i,1}(x) = \nabla B_i \cdot F(x)$, $d_2(x) = \nabla L \cdot F(x)$, 
		$d_{i,3} = 0.5 (d_{i,1}(x) + d_2(x)) + 0.5 |d_{i,1}(x) - d_2(x)|$, and sets
		$\mathcal{R}_{i,1}, \mathcal{R}_2, \mathcal{R}_{i,3}, \hat{\mathcal{R}}_{i,3}, i \in [1,N]$, are in \eqref{thmmulti:region}. 	
	\end{enumerate}
\end{theorem}

The proof of Theorem \ref{thmmulti:construct NCLBF} follows directly from Theorem \ref{thm3:construct NCLBF}, with regions
$\mathcal{R}_1$ and $\mathcal{R}_3$ replaced by $\mathcal{R}_{i,1}$ and $\mathcal{R}_{i,3}$, $i \in [1,N]$, respectively. The proof for each $\mathcal{R}_{i,1}$ and $\mathcal{R}_{i,3}$ is analogous. 
Note that the design constants $\eta_{i,1}$ and $\eta_{i,2}$, $i \in [1,N]$, must satisfy an additional condition  \eqref{thmmulti:eta2}. 
For each unsafe set $\mathcal{O}_i$, the proposed NCLBF \eqref{thmmulti:NCLBF} holds the properties similar to those in Section \ref{subsec:properties NCLBF}.
The points satisfying $x \in \mathcal{R}_{i,3}, ~ i \in [1,N]$, can be described as a sphere:
\begin{align}\label{sphere:Bx=Lx i}
	\| x - \bar{x}_{i,c} \|^2 = \bar{r}_i
\end{align}
with the centre $\bar{x}_{i,c} = \eta_{i,1} x_{i,c} / (1+\eta_{i,1})$ and radius $\sqrt{\bar{r}_i}$ satisfying $\bar{r}_i = [(1+\eta_{i,1}) \eta_{i,2} - \eta_{i,1} \| x_{i,c} \|^2 ] / (1+\eta_{i,1})^2$. Imposing the extra requirement in \eqref{thmmulti:eta2} ensures that the spheres for all unsafe sets remain disjoint.

Based on the NCLBF constructed in Theorem \ref{thmmulti:construct NCLBF}, the control design in Section \ref{subsec:control single} is extended to multiple unsafe sets, requiring Assumption \ref{assume:Pg multi}, analogous to Assumption \ref{assume:Pg}.

\begin{assume}\label{assume:Pg multi}
	For all $x \!\in\! \{ x \in \mathcal{R}_2 \cup (\cup_{i=1}^{N} \mathcal{R}_{i,3}) \mid \nabla L \cdot g(x) \!=\! 0  \}$, $\nabla L \cdot f(x) \!\leq\! 0$.	
	For all $x \!\in\! \{ x \in \mathcal{R}_{i,1} \cup \mathcal{R}_{i,3} \mid \nabla B_i \cdot g(x) \!=\! 0  \}$, $\nabla B_i \cdot f(x) \!\leq\! 0, i \in [1,N]$.	
	The system is zero-state detectable w.r.t. $\nabla L \cdot g(x)$ and $\nabla B_i \cdot g(x)$, i.e., $\nabla L \cdot g(x) = 0  ~\forall t \!\geq\! 0 \!\implies\! x(t) \rightarrow 0$ and $\nabla B_i \cdot g(x) = 0  ~\forall t \geq 0 \!\implies\! x(t) \rightarrow 0, i \in [1,N]$.
\end{assume}

The way of designing an effective controller to handle multiple unsafe regions is described in Theorem \ref{thm:control multi}.
\begin{theorem}\label{thm:control multi}
	Consider system \eqref{eq:sys1} under Assumption \ref{assume:Pg multi}, with the state space $\mathcal{X}$ and disjoint unsafe state sets $\mathcal{O}_i$, $i \in [1,N]$,	in \eqref{multi:obstacles}. By using the NCLBF in \eqref{thmmulti:NCLBF}, safe stabilization is achieved for $x(0) \in \mathcal{X} \setminus (\mathcal{O} \cup (\cup_{i=1}^{N} \mathcal{R}_{i,1}) \cup (\cup_{i=1}^{N} \mathcal{R}_{i,3}))$, where $\mathcal{O} = \cup_{i=1}^N \mathcal{O}_i$, if the controller is designed as
	\begin{align}\label{controller: multi}
		u = 
		\begin{cases}
			\kappa_{i,1}(x), &  x \in \mathcal{R}_{i,1}, ~ i \in [1,N] \\
			\kappa_2(x), & x \in \mathcal{R}_2 \\
			\kappa_{i,3}(x), & x \in \mathcal{R}_{i,3}, ~ i \in [1,N]
		\end{cases}
	\end{align}
	with
	\begin{subequations}\label{controllers: multi}
		\begin{align}
			\kappa_{i,1}(x) &= 
			\begin{cases}
				-\mu(B_{i,g}) B_{i,f} - \mathbf{c}_{i,1} \bar{\mu}(B_{i,g}) \| x \|^2, \!\!&\!\! B_{i,g} \neq 0 \\
				0, \!\!&\!\! B_{i,g} = 0 \!\!
			\end{cases},\!\!\!\!
			\nonumber\\
			\kappa_2(x) &= 
			\begin{cases}
				- (L_f + \sqrt{L_f^2 \!+\! \gamma \| L_g \|^4}) \mu(L_g), & L_g \neq 0 \\
				0, & L_g = 0	
			\end{cases}, \nonumber\\
			\kappa_{i,3}(x) &= 
			\begin{cases}
				\kappa_{i,1}(x), \!\!&\! (x(t) \in \mathcal{R}_{i,3}) \cap (x(t-t_s) \in \mathcal{R}_{i,1}) \\
				\kappa_2(x), \!\!&\! (x(t) \in \mathcal{R}_{i,3}) \cap (x(t-t_s) \in \mathcal{R}_2) \\
				\kappa_2(x), \!\!&\! (x(t) \in \mathcal{R}_{i,3}) \cap (x(t-t_s) \in \mathcal{R}_{i,3})
			\end{cases}, \nonumber
		\end{align}	
	\end{subequations}
	with $B_{i,f} = \nabla B_i \cdot f(x)$, $B_{i,g} = \nabla B_i \cdot g(x)$, $L_f= \nabla L \cdot f(x)$, $L_g = \nabla L \cdot g(x)$, the design constants
	$\mathbf{c}_{i,1} = \mathrm{diag}(c_{i,1,1}, \cdots, c_{i,1,j}) > 0, j \in [1,m]$, $\gamma > 0$, and $\eta_{i,1}$ and $\eta_{i,2}$ satisfying \eqref{thmmulti:eta}. 
	The sets $\mathcal{R}_{i,1}, \mathcal{R}_2$ and $\mathcal{R}_{i,3}, i \in [1,N]$, are defined in \eqref{thmmulti:region}.
\end{theorem}

The proof of Theorem \ref{thm:control multi} follows directly from Theorem \ref{thm:control design} and is therefore omitted. As shown in \cite{braun2017existence,braun2019complete}, the continuous universal control law based on smooth CLBF \cite{romdlony2016stabilization} can prevent the state from entering unsafe regions but may introduce undesired local equilibria due to gradient cancellation. In this paper, we address this issue by introducing the NCLBF with $\max$ operation, ensuring control design effectiveness compared to CLBF-based methods. A performance comparison of both approaches is provided in Section \ref{sec:example}

\section{Illustrative examples}\label{sec:example}

\subsection{Linear System with A Single Unsafe Set}\label{subsec:sim1}
Consider the linear system \cite[Example 14]{braun2017existence}:
\begin{align}\label{eg1:sys}
	\begin{split}
		\dot{x}_1 &= - x_1 + u_1 \\
		\dot{x}_2 &= - x_2 + u_2
	\end{split}
\end{align}
with the state space $\mathcal{X} := [-5,5] \times [-5,5]$.
Rewriting \eqref{eg1:sys} as \eqref{eq:sys1} with $x = [x_1, x_2]^\top$, $u = [u_1, u_2]^\top$, $f(x) = -x$, and $g(x) = I_2$.
The goal is steering $x$ to the origin while avoiding unsafe states in $\mathcal{D} := (1,3) \times (1,3)$.
The set $\mathcal{D}$ can be enclosed by  $\mathcal{O} = \{ x \in \mathcal{X} \setminus \{0\} \mid \| x - x_c \| < \sqrt{2} \}$, with $x_c = [2,2]^\top$.  

We apply the NCLBF and control design from Section \ref{sec:single}. Note that $\nabla L \cdot g(x) = 2 x^\top = 0$ only at $x = 0$, while $\nabla L \cdot f(x) = -2 x^\top x \leq 0$. Further noting that $\nabla B \cdot g(x) = - 2 \eta_1 (x - x_c)^\top = 0$ only at $x = x_c \in \mathcal{O}$. Moreover, when $x = 0$ or $x = x_c$ for all $t \geq 0$, $x(t) \rightarrow 0$ (even with $u(t) = 0$). 
Hence, Assumption \ref{assume:Pg} holds.
The design parameters of our method are: $\eta_1 = 9$, $w = 0.9$, $\eta_2 = 36.9$, $\mathbf{c}_1 = \mathrm{diag}(10,20)$, $\gamma = 0.1$. A comparison is made with the CLBF method \cite{romdlony2016stabilization} by borrowing the design parameters from \cite[Example 14]{braun2017existence}.

\begin{figure}[t]
	\centering
	\includegraphics[width=\linewidth]{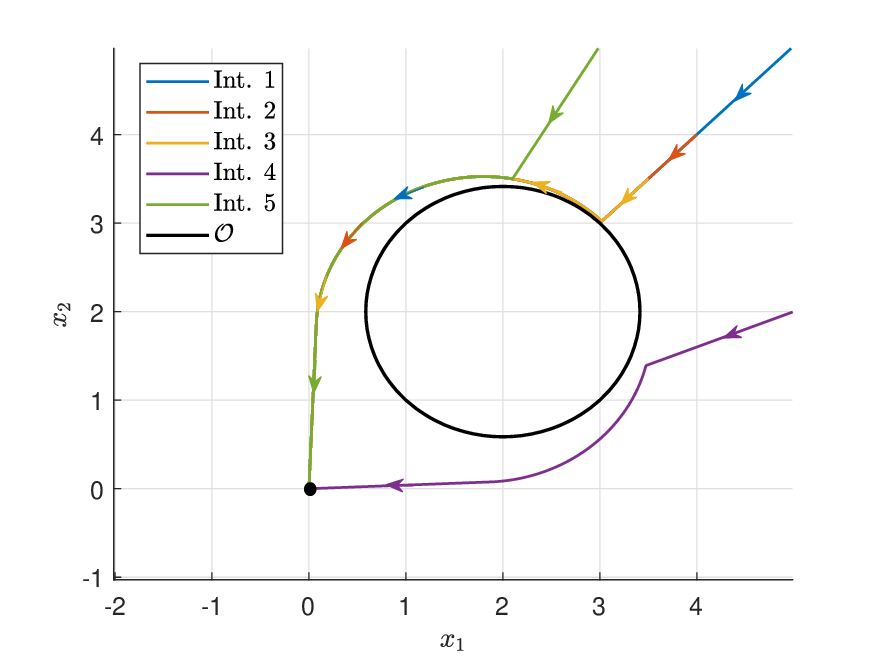}
	\vspace{-6mm}
	\caption{Trajectories for five initial states: Section \ref{subsec:sim1}, NCLBF.}
	\label{fig4}
\end{figure}

\begin{figure}[t]
	\centering
	\includegraphics[width=\linewidth]{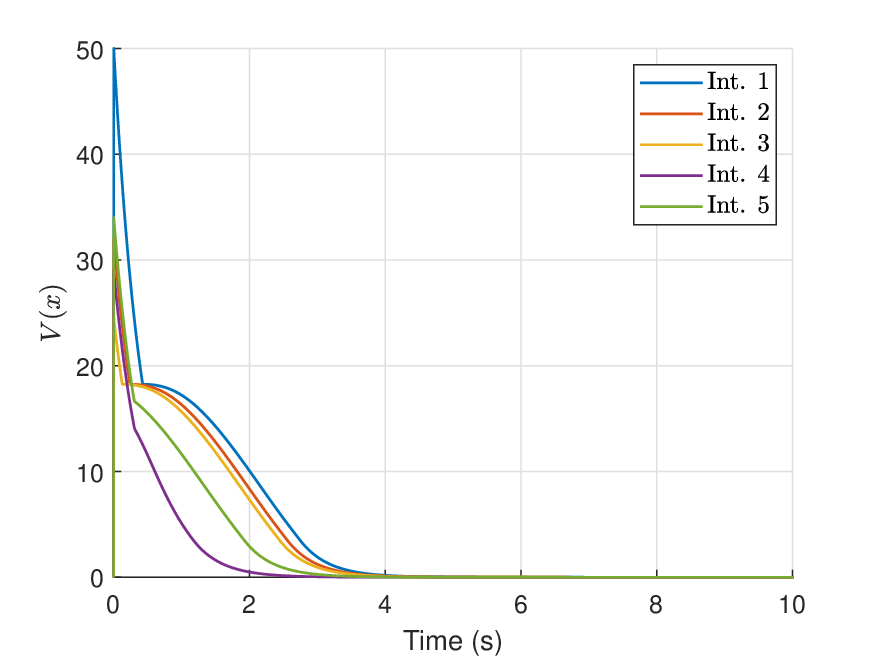}
	\vspace{-6mm}
	\caption{$V(x)$ for five initial states: Section \ref{subsec:sim1}, NCLBF.}
	\label{fig5}
\end{figure}

\begin{figure}[t]
	\centering
	\includegraphics[width=\linewidth]{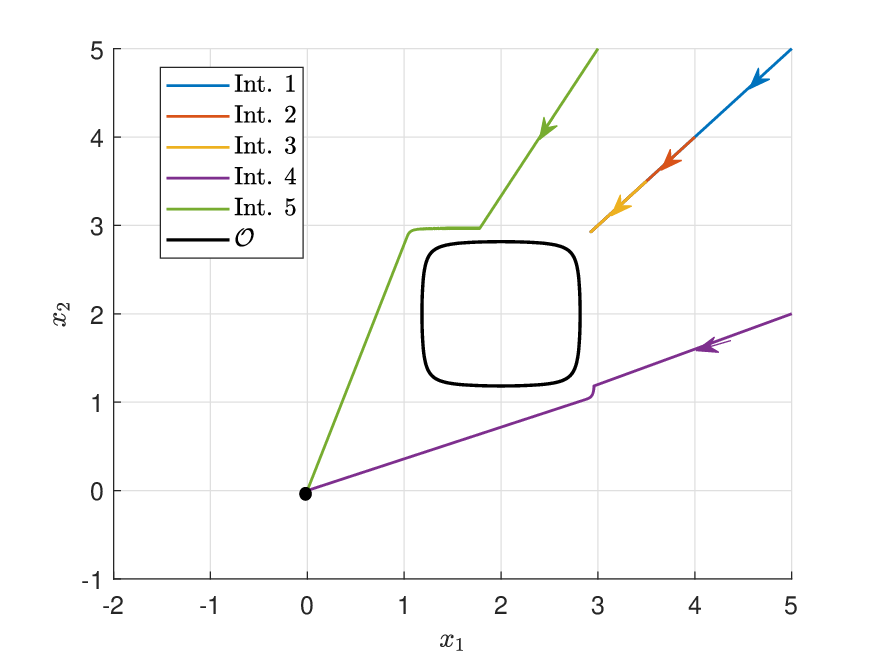}
	\vspace{-6mm}
	\caption{Trajectories for five initial states: Section \ref{subsec:sim1}, CLBF.}
	\label{fig7}
\end{figure}

Both controllers are tested on five initial (Int.) states $x(0)$: $(5,5), (4,4), (3.5,3.5), (5,2), (3,5)$.
As shown in Fig. \ref{fig4}, the proposed controller successfully guides the state to the origin while avoiding the unsafe set in all cases. Moreover, for each initial state, $V(x)$ decreases over time, as illustrated in Fig. \ref{fig5}.
However, as shown in Fig. \ref{fig7}, the CLBF-based controller fails to steer the state to the origin when the initial condition satisfies 
$x_1(0) = x_2(0) > 3$.

\begin{figure}[t]
	\vspace{-4mm}
	\centering
	\includegraphics[width=\linewidth]{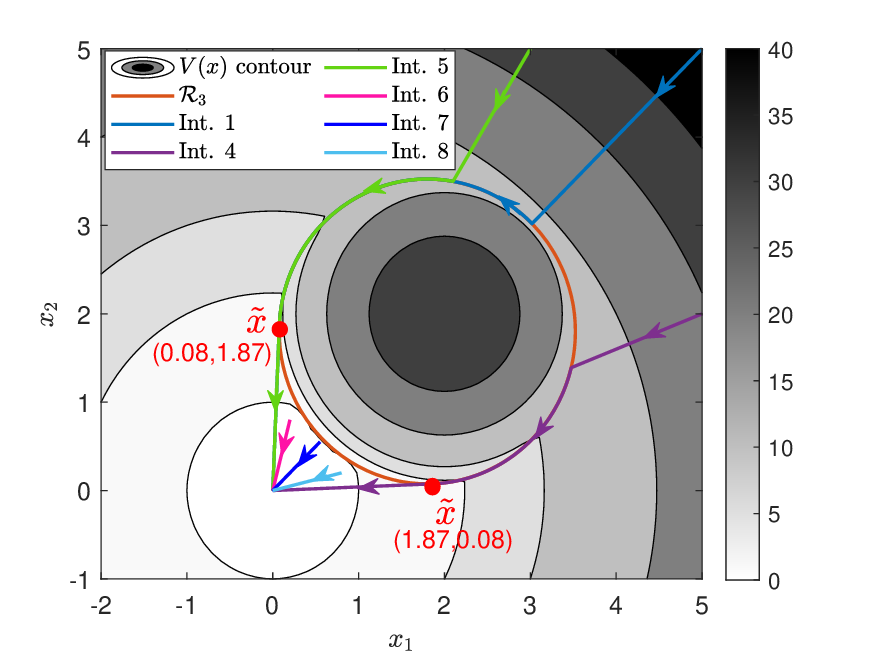}
	\vspace{-6mm}
	\caption{Trajectories for eight initial states illustrating Proposition \ref{prop:NCLBF reduce R3}: Section \ref{subsec:sim1}, NCLBF.}
	\label{fig6}
\end{figure}

We further validate the proposed approach of shrinking $\mathcal{R}_3$ in Proposition \ref{prop:NCLBF reduce R3}, by analyzing the state trajectories and corresponding $V(x)$ values for Int. 1, Int. 4 and Int. 5, along with three extra initial states: Int. 6 $(0.2, 0.8)$, Int. 7 $(0.55, 0.55)$ and Int. 8 $(0.8, 0.2)$, as shown in Fig. \ref{fig6}. 

According to Proposition \ref{prop:NCLBF reduce R3}, we compute $\phi(x_c) = 3.50$. The intersection points, i.e. the contacting points of the tangent planes between $\|x\|^2 = \phi(x_c)$ and $\mathcal{R}_3$, are determined as $\tilde{x} = (1.87,0.08)$ and $\tilde{x} = (0.08,1.87)$. As depicted in Fig. \ref{fig6}, the contacting point $\tilde{x} = (1.87,0.08)$ aligns precisely with the actual point where the state trajectory of Int. 4 exits $\mathcal{R}_3$ and moves directly toward the origin. 
Likewise, $\tilde{x} = (0.08,1.87)$ corresponds to the points where the state trajectories of Int. 1 and Int. 5 leave $\mathcal{R}_3$ and converge to the origin.
The results in Fig. \ref{fig6} confirm that the state trajectories do not enter $\hat{\mathcal{R}}_3$ from above (Int. 1, Int. 4 and Int. 5) or below (Int. 6, Int. 7 and Int. 8), instead converging to the origin. It thus validates the soundness of Proposition \ref{prop:NCLBF reduce R3}.

\subsection{Nonlinear System with Multiple Unsafe Sets}\label{subsec:sim3}
We consider the nonlinear mechanical system:
\begin{align}\label{case4:sys}
	\begin{split}
		\dot{x}_1 &= x_2 \\
		\dot{x}_2 &=  - x_1 - x_2 - (0.8 + 0.2 \mathrm{e}^{-100 |x_2|} ) \tanh(10 x_2) + u	\!\!
	\end{split}	
\end{align}
with the state space $\mathcal{X} := [-5,5] \times [-5,5]$ and three unsafe state sets 
$\mathcal{D}_1 := (1.2,2.8) \times (-0.8,0.8)$,  
$\mathcal{D}_2 := (1.3,2.7) \times (1.3,2.7)$, and $\mathcal{D}_3 := (-3,-1) \times (-1,1)$. These sets are enclosed by $\mathcal{O}_1 = \{ x \in \mathcal{X} \setminus \{0\} \mid \| x - x_{1,c} \| < \sqrt{0.7} \}$, $\mathcal{O}_2 = \{ x \in \mathcal{X} \setminus \{0\} \mid \| x - x_{2,c} \| < \sqrt{0.5} \}$,  and $\mathcal{O}_3 = \{ x \in \mathcal{X} \setminus \{0\} \mid \| x - x_{3,c} \| < 1 \}$ with $x_{1,c} = [2,0]^\top$, $x_{2,c} = [2,2]^\top$ and $x_{3,c} = [-2,0]^\top$, respectively. 

We apply the NCLBF and control design from Section \ref{sec:multiple}, first verifying Assumption \ref{assume:Pg multi}.
Through trivial calculation, we have the following: 
1) $\nabla L \cdot g(x) = 2 x_2 = 0$ only at $x_2 = 0$, while $\nabla L \cdot f(x) = 0$;
2) $\nabla B_1 \cdot g(x) =  - 2 \eta_{1,1} x_2 = 0$ only at $x_2 = 0$, while $\nabla B_1 \cdot f(x) = 0$;
3) $\nabla B_2 \cdot g(x) =  - 2 \eta_{2,1} (x_2 - 2) = 0$ only at $x_2 = 2$, while $\nabla B_2 \cdot f(x) = 4 \eta_{2,1} (2 - x_1)$;
4) $\nabla B_3 \cdot g(x) =   - 2 \eta_{3,1} x_2 = 0$ only at $x_2 = 0$, while $\nabla B_3 \cdot f(x) = 0$.
We further examine the case 3) when $\nabla B_2 \cdot g(x) = 0$.  
Note that $0 < x_1 < 2 - \sqrt{0.5}$ or $x_1 > 2 + \sqrt{0.5}$ in $\mathcal{R}_{2,1} \cup \mathcal{R}_{2,3}$. If $x_1 > 2 + \sqrt{0.5}$, then $\nabla B_2 \cdot f(x) < 0$. If at time $t$, $x_2(t) = 2$ and $0 < x_1(t) < 2 - \sqrt{0.5}$, then $\dot{x}_1 = 2, \dot{x}_2 = - x_1(t) - 2.8 + u(t)$, ensuring $x_1$ increases and $x_2$ decreases (even when $u(t) = 0$), leading to $\nabla B_2 \cdot g(x) \neq 0$ at time $t + 1$.
In summary, Assumption \ref{assume:Pg multi} is satisfied. 

The following parameters are used to design the NCLBF-based controller: 
$\eta_{1,1} = 11$, $w_1 = 0.3$, $\eta_{1,2} = 16$, $\mathbf{c}_{1,1} = 10$,
$\eta_{2,1} = 19$, $w_2 = 0.7$, $\eta_{2,2} = 22.7$, $\mathbf{c}_{2,1} = 20$,
$\eta_{3,1} = 18$, $w_3 = 0.2$, $\eta_{3,2} = 27.2$, $\mathbf{c}_{3,1} = 20$,
and $\gamma = 5$.

\begin{figure}[t]
	\centering
	\includegraphics[width=\linewidth]{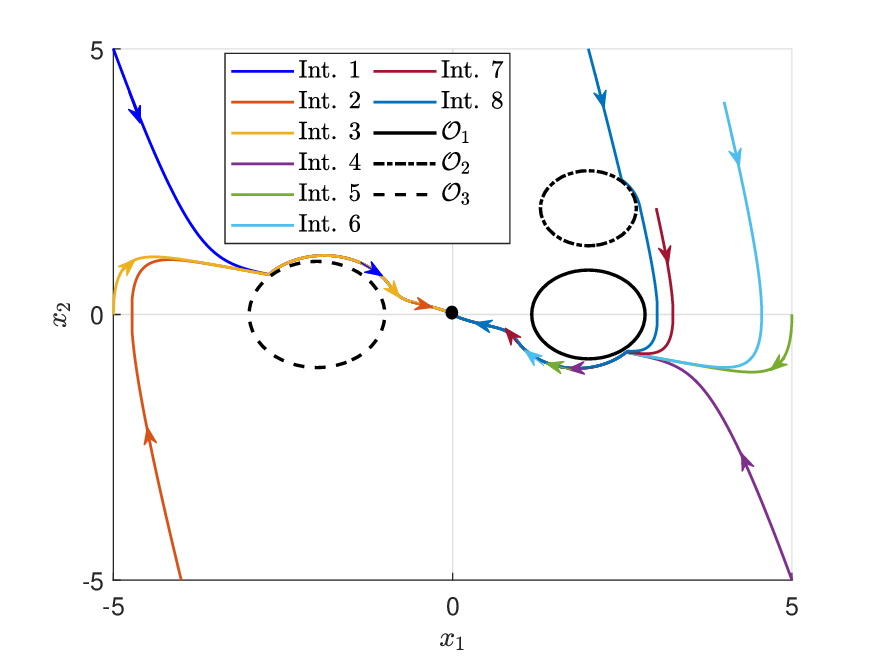}
	\vspace{-6mm}
	\caption{Trajectories for eight initial states: Section \ref{subsec:sim3}.}
	\label{fig13}
\end{figure}

\begin{figure}[t]
	\centering
	\includegraphics[width=\linewidth]{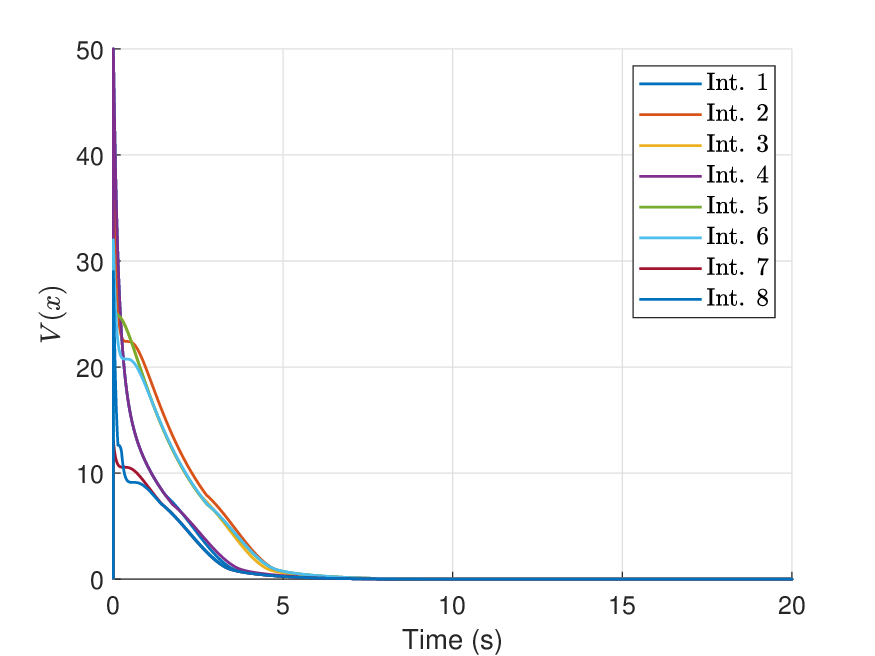}
	\vspace{-6mm}
	\caption{$V(x)$ for eight initial states: Section \ref{subsec:sim3}.}
	\label{fig14}
\end{figure}

The proposed controller is simulated for eight different initial (Int.) states $x(0): (-5,5)$, $(-4,-5)$, $(-5,0)$, $(5,-5)$, $(5,0)$, $(4,4)$, $(3,2)$, $(2,5)$.  
As shown in Figs. \ref{fig13} and \ref{fig14}, the controller successfully guides the state trajectories to the origin while avoiding unsafe sets for all initial conditions.

\section{Conclusions}\label{sec:conclusion}
This paper presents the nonsmooth Lyapunov barrier function (NCLBF) for safe stabilization in systems with multiple bounded unsafe sets. By leveraging a systematic design of NCLBF and piecewise continuous feedback control, it overcomes the limitations of smooth Lyapunov barrier methods. The approach is validated through rigorous theoretical analysis and extensive simulations. Future work will explore extending it to more complex operational scenarios using Boolean logic for state constraints.

\bibliographystyle{IEEEtran} 
\bibliography{reference}

\end{document}